# Huge Volume Expansion and Structural Transformation of Carbon Nanotube Aligned Arrays during Electrical Breakdown in Vacuum

Shashank Shekhar [1,2], Helge Heinrich [2,3], and Saiful I. Khondaker [1,2,4]*

[1] Nanoscience Technology Center, [2] Department of Physics, [3]Advanced Material Processing and Analysis Center, [4]School of Electrical Engineering and Computer Science, University of Central Florida, Orlando, Florida 32826, USA.

**Abstract**

We observed a huge volume expansion of aligned single walled carbon nanotube (*SWNT*) arrays accompanied by structural transformation during electrical breakdown in vacuum. The *SWNT* arrays were assembled between prefabricated *Pd* source and drain electrodes of 2 *μm* separation on *Si/SiO₂* substrate via dielectrophoresis. At high electrical field, the *SWNT* arrays erupt into large mushroom-like structure. Systematic studies with controlled electrical bias show that above a certain field *SWNTs* swell and transform to nanoparticles and flower-like structures with small volume increase. Further increase in electrical bias and repeated sweeping results into amorphous carbon as determined from scanning and transmission electron microscopy (*TEM*). Cross sectional studies using focused ion beam and *TEM* show the height of 2-3 *nm* *SWNT* array increased to about 1 *μm* with a volume gain of ~ 400 times. The electron energy loss spectroscopy reveals that graphitic $sp^2$ networks of *SWNTs* are transformed predominantly to $sp^3$. The current-voltage measurements also show an increase in the resistance of the transformed structure.

## 1. Introduction

Single walled carbon nanotubes (*SWNTs*) exhibit many noble properties such as ballistic transport, high mobility, high thermal conductivity, strong mechanical strength and high current carrying density ($10^9$ A/cm$^2$).[1-15] *SWNTs* have also exhibited many interesting quantum behaviors such as universal conductance fluctuation, quantum conductance, resonant tunneling, single electron tunneling and spin polarized conductance.[4,12-15] Due to its geometry, *SWNT* demonstrates promising capillary action and field emission properties.[16-17] In extreme conditions carbon nanotubes have shown many fascinating behaviors. At high temperatures, carbon nanotube exhibits superplasticity and elongates to 280% under application of stress.[18] In liquid crystal medium, the cluster of carbon nanotubes shows super-elongation of 400% due to electroactive action which is restored to original length after removal of field.[19] At very high temperatures (~1500 $^0$C), individual *SWNTs* exhibit coalescence with neighboring nanotubes due to bond breakage that results in diameter enhancement of nanotubes.[20-21] Under the influence of external energy such as laser irradiation, electron irradiation, and ion beam irradiation graphitic networks reorganize their structures and the intriguing behaviors in nanotubes are observed due to defect creation.[22] In such conditions carbon nanotubes are transformed into various fascinating structures such as carbon onion, and nano-diamond.[22]

Recently, aligned arrays of *SWNT* have emerged as an important class of material due to their superior properties and multifunctionality compared to individual nanotubes.[23-40] Aligned arrays of nanotubes sheet have been found to be structurally homogeneous, highly conducting, optically transparent and flexible. Devices fabricated from arrays of *SWNTs* can be advantageous as they average out device to device inhomogeneity of individual *SWNTs* and cover large areas. For these reasons, there is a significant research effort to fabricate devices with massively parallel arrays of *SWNTs* for various applications.[23-40] Recently we have shown an intriguing new property of correlated electrical breakdown of nanotubes in air in a densely aligned array resulting in formation of nano-fissure like shape.[41] This phenomenon arises from the fact that the aligned arrays of *SWNTs* are not just merely a collection of nanotubes but behave in a more complex way and they are capable of demonstrating a collective behavior.[41]

In this study we report yet another interesting phenomenon of aligned arrays of *SWNTs* under high electrical bias in vacuum. We show that densely aligned *SWNT* array undergoes a huge volume expansion along with a structural transformation from $sp^2$ graphitic network to predominantly $sp^3$ amorphous carbon (*a-C*) structures during electrical breakdown in vacuum on a *Si/SiO₂* substrate. The overall volume gain of the aligned array was estimated to be ~ 400 times. Detailed systematic studies show that, during the electrical breakdown *SWNTs* first swell and coalesce due to bond breakage and their diameter increases. Repeated electrical sweeps result in mushroom like structure consisting of *a-C* as determined from scanning electron microscopy (*SEM*) and transmission electron microscopy (*TEM*) analysis. Cross sectional analysis using focused ion beam (*FIB*) and *TEM* show that the transformed structure is predominantly carbon with a lots of voids and there is only small substrate damage. Electron energy-loss spectroscopy (*EELS*) also reveals that the mushroomed structure consists of amorphous carbon with predominantly $sp^3$ bonding and confirms bond rearrangement during vacuum electrical breakdown. The low bias current-voltage characteristics show that the resistance of the network increased due to the structural transformation.

## 2. Experimental

**Device Fabrication.** The devices were fabricated on highly doped *Si* substrates with a 250 *nm* thick *SiO₂* caped layer. Larger electrode patterns and contact pads were fabricated by standard optical lithography followed by thermal deposition of 3 *nm Cr* and 50 *nm* thick *Au*. Smaller source and drain electrode patterns were defined with electron beam lithography (*EBL*) and electron beam deposition of 2 *nm Cr* and 25 *nm* thick *Pd* followed by lift-off. The length of the channel was 2 *μm* while the width of the channel was 25 *μm*. The directed assembly of *SWNTs* at predefined electrodes was done in a probe station under ambient conditions via dielectrophoresis. A stable, surfactant free and catalytic particle free solution of *SWNT* dispersed in water was obtained from Brewer Science that had a *SWNT* concentration of ~50 μg/ml.[42] After dilution of the solution using de-ionized water to 4 ~ *μg/ml* concentrations, a 3*μL* drop was cast onto the electrodes. We applied an *AC* voltage of 5 *Vp-p* at 300 *kHz* between the drain and source electrodes for 30 seconds. Finally the surface was blown dry with nitrogen gas.

**Electrical Measurements and Characterizations**. The chip was wire bonded and mounted on a chip holder. The electrical measurements were performed inside a cryostat having 30 mTorr of vacuum. The electrical breakdown was performed with the help of Keithely 2400 sourcemeter and 6517 electrometer interfaced with LabView. The devices were imaged using a Zeiss Ultra 55 field emission scanning electron microscope. The microscope is capable of delivering high lateral resolution at low voltages. Inlens, and Secondary electron detectors were used to get images at low voltages. *AFM* images were collected in tapping mode by the Veeco manifold multimode and dimension 3100 instruments. The scan rate of the *AFM* was 1 *μm/sec* and 256×512 pixels were scanned for each image. For the *TEM* analysis samples were prepared by *FIB* milling with instrument FEI 200 TEM FIB. First 2-3 *nm* of *Au-Pd* is sputter deposited on the specimen by sputtering. Then 1 *μm* Pt metal was deposited by ion beam assisted chemical vapor deposition inside the *FIB* instrument. *FIB* instrument removes material around specimen by sputtering using gallium at lateral resolution of approximately 5 *nm*. The cross section of the specimen was collected for *TEM* analysis. The *HRTEM* and *EELS* analysis on the sample was performed by Technai F30 FEI transmission electron microscope. The microscope is equipped with a cold field emission source, and a Gatan imaging energy loss spectrometer.

### 3. Results and Discussion

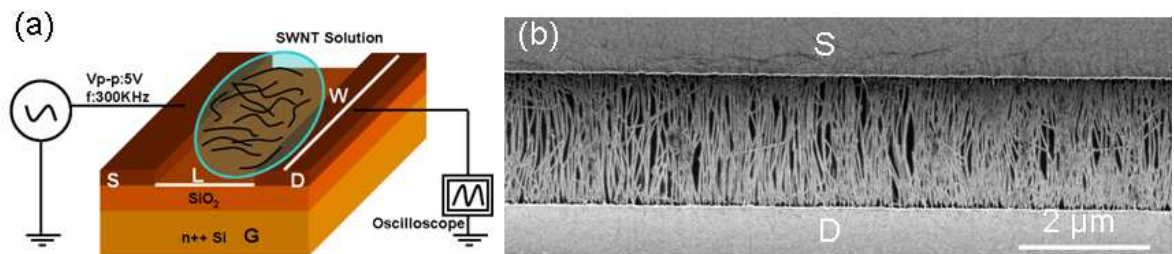

Figure 1. Dielectrophoretic assembly of *SWNT* array. (a) Schematic of device geometry. *SWNTs* are aligned on *Pd* electrodes of channel length 25 *μm* and width 2 *μm*. (b) *SEM* image of a dielectrophoretically assembled array of *SWNTs* with linear density ~30/*μm*.

Figure 1a shows the schematic diagram of the assembly set up on a conventional $Si/SiO_2$ substrate having an oxide thickness of 250 *nm*. The channel length (*L*) of the device was 2 *μm* whereas channel width (*W*) was 25 *μm*. *SWNTs* were assembled between *Pd* drain (*D*) and source (*S*) electrodes by dielectrophoresis from an aqueous solution of *SWNT*. The details of the assembly have been discussed in our previous publication.[25] In brief, a small (3 *μL*) drop of the *SWNT* solution was cast onto the chip containing the electrode arrays. An *AC* voltage of 5 Vp-p with a frequency of 300 *kHz* was applied using a function generator between the pair of *S* and *D* electrodes for 30 s. Figure 1b shows the *SEM* image of a typical device of *SWNT* array consisting of 30 *SWNT/μm*. The sample was then bonded and loaded into a vacuum cryostat for electrical measurements.

Figure 2a shows successive current (*I*) – electric field (*E*) characteristics of a representative *SWNT* array measured in vacuum of 30 *mTorr*. During the first sweep, current initially increases rapidly to a few milli-Amperes as *E* is increased and then sharply decreases for *E* > 6.0 *V/μm*. However, the current remains in the *mA* range and then starts to increase slowly again after 10 *V/μm*. It is interesting to note that current never comes down to zero, as expected from electrical breakdown of individual carbon nanotubes.[9] Subsequent electric field sweep on the same sample surprisingly show similar trend with first sweep. *SEM* image (figure 2b) taken after the electrical sweeping shows huge mushroom-like structure. It appears that the *SWNT* between the electrodes erupted with no visible sign of any *SWNT* left in the circuit. This observed phenomenon is in sharp contrast with the electrical breakdown of the array in air where nano-fissure like structure was observed. This is shown in figure 2c (*I-E* curve) and figure 2d (*SEM* image). The details of the nano-fissure formation in air can be found in our previous

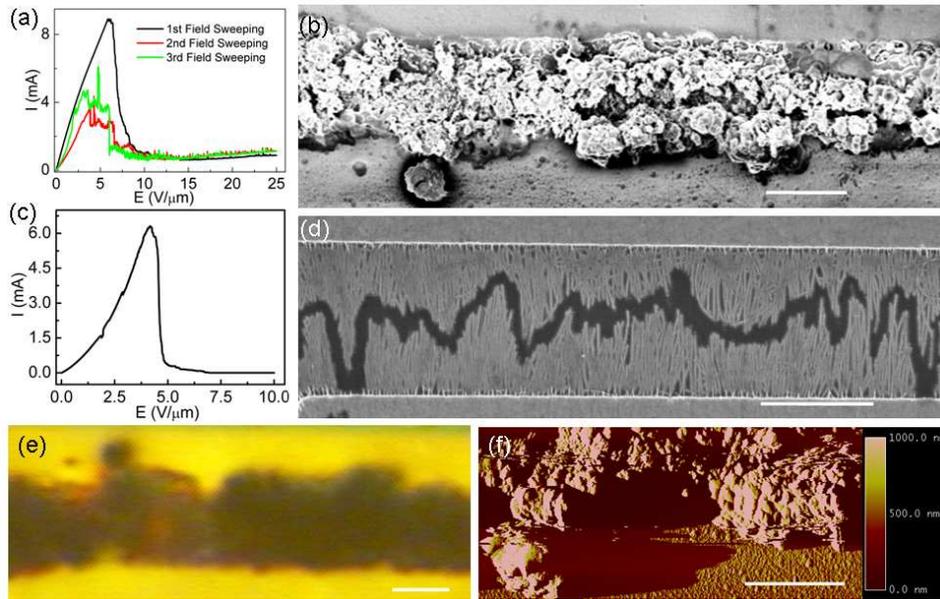

Figure 2. Electric field sweeping of SWNT array and structural transformation. (a) Current (*I*)- electric field (*E*) characteristic of SWNT array in vacuum for three successive sweeps. The current does not become zero even at 25 *V/µm* (b) SEM of the structure after the field sweep. Huge mushroom like structure can be seen. (c) *I - E* characteristics of the array in air. Above 4 *V/µm I* falls rapidly and no current is measured above 10 *V/µm*. (d) *SEM* shows that array of *SWNTs* undergo electrical breakdown in air giving rise to nano-fissure like structure. (e) Optical image of the vacuum broken sample shows black and charred material between the electrodes. (f) *AFM* image of the aligned array after vacuum field sweep. The height is ~ 1 *µm*. Scale bars are report.[41] Figure 2e shows the optical image of the new structure formed during the vacuum breakdown. Unlike the optical image of the aligned array which is transparent (see supporting information figure S2), this image shows the formation of an opaque and black material between the electrodes. The optical image suggests that the new material could be due to a structural transformation. There is also a possibility that electrodes or substrate got damaged during the sweeping due to Joule heating. We have also performed atomic force (*AFM*) microscopy measurement on the transformed material. This is shown in figure 2f where we see the height of the structure is ~ 1*µm*. For more accurate height determination, we have done cross sectional analysis which we have discussed later. The *SEM* and *AFM* images clearly show that the *SWNT* array has transformed to a new structure with a huge volume gain.

In order to investigate the new materials in more details, we have performed high resolution *SEM* (*HRSEM*) and *TEM* analysis. Figure 3a & 3b show *HRSEM* image of the material from two different regions showing varying topology of the materials. Spherical structures that are connected to each other possibly due to reconstruction can be seen in the figure 3a & 3b. The size of these structures are estimated to vary from 20 -200 *nm*. Figure 3c shows high magnification image for part of figure 3b where particles of 20 *nm* diameter is clearly seen. These micrographs show no trace of individual *SWNT*. In addition, a lot of voids and pores can also be seen. For *TEM* measurements, the material was taken out by scooping with a micro-needle in a micromanipulator and put on the lacey carbon grid. Figure 3d shows a *TEM* image of the scooped material. In high-resolution *TEM* (*HRTEM*) image (figure 3e) of the materials we did not find any crystalline pattern and the structure appears to be amorphous. Figure 3f shows the selected area electron diffraction (*SAED*) pattern with one broad diffuse ring confirming the amorphous structure. It is known that *SWNT* possesses crystalline graphitic sp$^2$ network.[43-44]

Therefore, the *TEM* analysis reveals that a structural transformation takes place during the vacuum breakdown that converts the crystalline structure to amorphous material.

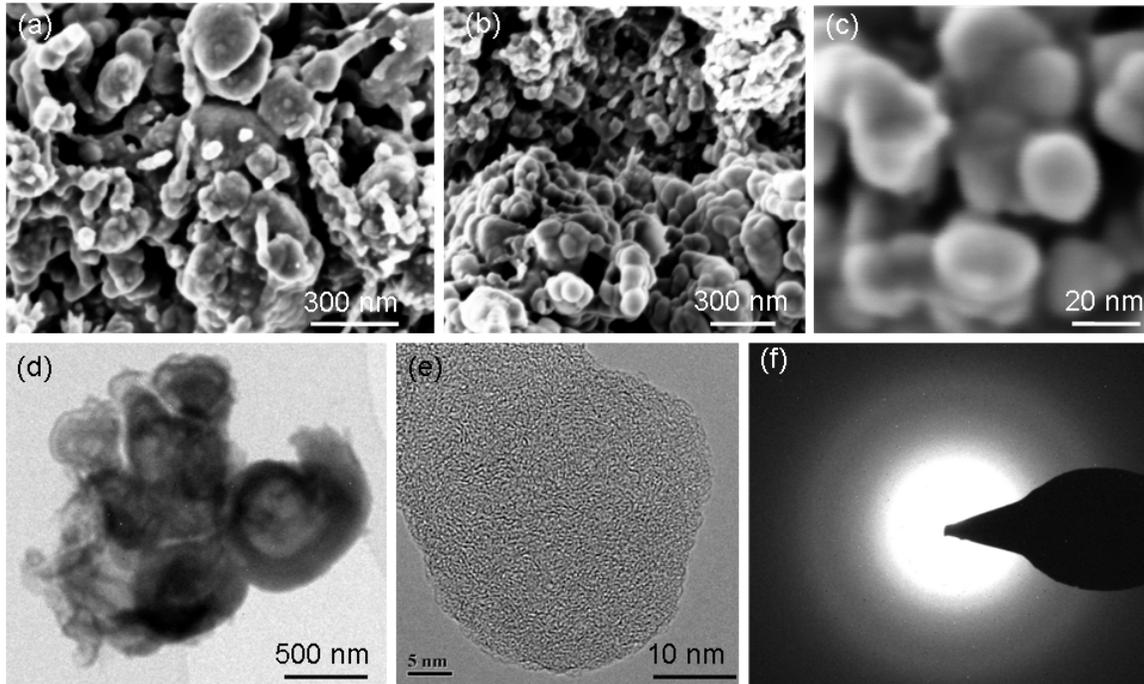

Figure 3. Microstructure of the transformed material. (a-c) *SEM* images of the transformed material. A wide range of structures of size 20 -200 *nm* are visible. The particles are melted and fused together. Lots of voids can also be seen. (d) A *TEM* image of the material (e) *HRTEM* of the part of a particle shows amorphous structure (f) Diffused electron diffraction ring of *SAED* confirms amorphous nature.

In order to understand how the *SWNT* got transformed into this mushroomed structure, we have done controlled electrical breakdown by stopping the sweep at $E = 10$ *V/μm* as shown in Figure 4a. The *SEM* image subsequent to this controlled experiment is presented in figure 4b. Several interesting phenomena are observed. First, the portion of the nanotubes closer to the source electrode (positive bias) is swollen and coalesces with each other at some places while the portion of the nanotubes closer to the drain remains intact and thus forms structural discontinuity. Second, small particles are formed at the discontinuity points bridging the two portions of the nanotubes. The average size of these particles is ~ 10 *nm*. Third, interesting flower like patterns can be seen at the tip of the swollen nanotubes near the structural discontinuity. These interesting structures are more clearly seen in the *HRSEM* images of figure 4c, d, e and f. We have done *AFM* studies on these structures (see supporting information figure S3). From the *AFM* image, we found the diameter of the swollen tubes to vary from 20-30 *nm*. The diameter of the flower like structure lies between 100-200 nm. It appears that during the electrical breakdown process nanotubes first swell to a tubular structure of larger diameter. This is possible with restructuring and reorganization of bonds. As the field is increased further the swollen tubes start to coalesce with each other and further gain the volume. The structure becomes unstable and disintegrates into particles of various shapes and sizes. The high electric field and joule heating provide the required energy for restructuring the bonds.

Recently, it has been reported that an individual *SWNT* leaves a thermal signature on *Si/SiO₂* substrate during the vacuum breakdown although no swelling of *SWNT* was observed.[45] Therefore, a very important question arises that whether the transformed structure consists of solely carbon or the underlying substrate also get damaged and contributes in the mushroomed structure. In order to investigate the substrate effect and get more insight into the structure, we performed cross-sectional analysis through *FIB* cutting of the sample and subsequent *TEM* imaging. Figure 5a shows *FIB* image of one of our samples before the *FIB* milling. For cross sectional analysis a small part of the sample was cut by milling with *Ga* ion. Two specimens were prepared from two different samples. Both samples were prepared from the middle portion, one is along the channel (specimen *A*) while the other was orthogonal to the channel (specimen *B*). Figure 5b shows the *FIB* image (tilted at 45°) of specimen *A* while 5e show the image of specimen *B* after *FIB* milling. The samples were then cut and placed on the *TEM* grid for *TEM* analysis.

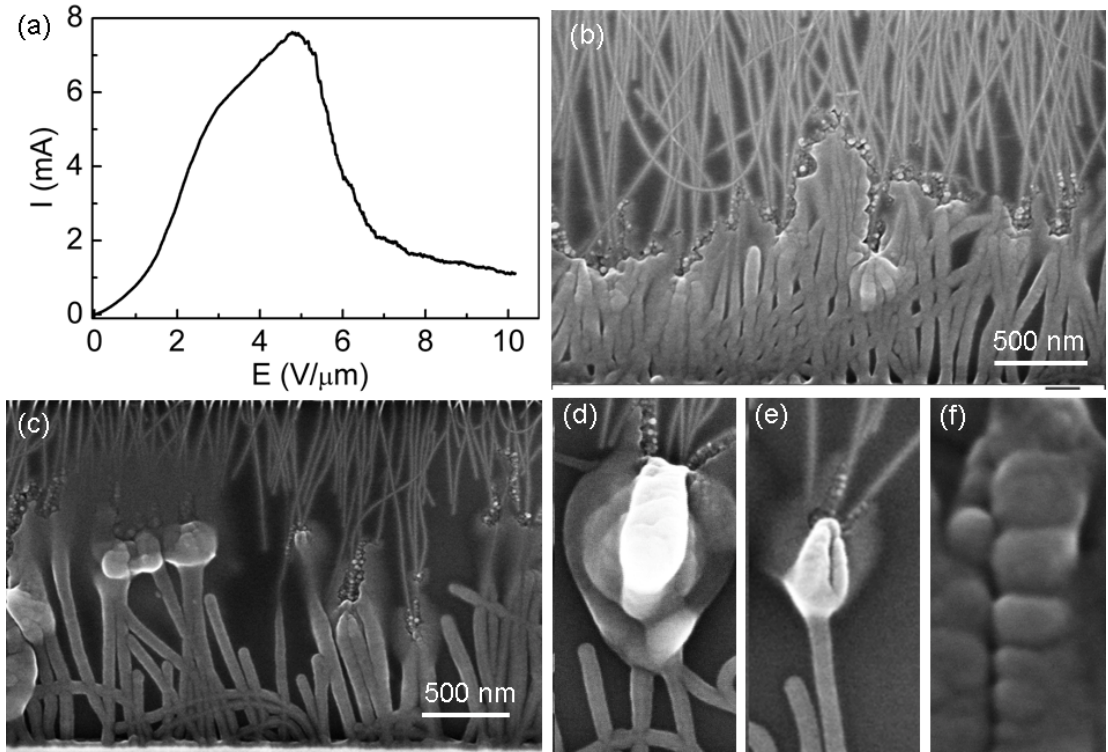

Figure 4. Transformation of the *SWNT* array during controlled field sweeping. (a) The field sweeping was restricted up to 10 *V/μm*. (b) *SEM* image after the controlled sweep show part of the SWNT remains intact while other parts are swollen and coalesced. Small nanoparticles can be seen at the structural discontinuity point. (c)-(f) structural diversity in the SWNT array during controlled sweep.

Figure 5c is a low resolution *TEM* image of specimen *A* showing the cross section. Huge volume expansion in the structure is confirmed by the cross sectional analysis. A lot of voids and pores (bright spots in the image) are seen within the structure consistent with *SEM* observations. The initial thickness of aligned array was ~ 2-3 *nm*, and after mushrooming it becomes ~1000 *nm*. Hence the volume of the structure is enhanced by ~ 400 times. In figure 5d the interface between 250 *nm* thick $SiO_2$ substrate and transformed structure is clearly visible. A minor dent to the substrate is seen at some places though we did not observe $SiO_2$ diffusion into the structure. A larger void is also seen in this image. We also examined specimen *B* cut orthogonal to the substrate as shown in figure 4e. This specimen contains the cross section of *Pd* electrodes and can provide the evidence of any possible damage to them. *HRTEM* figure 4f shows a similar phenomenon of volume expansion. Small voids are also seen in the structure (bright spots). The black line above the substrate is the *Pd* electrode which appears undisturbed. Figure 5g is TEM of the selected area marked in figure 5f. In this particular cross section also, we observe minor damage to the substrate at few places. We also find diffusion of *Pt* metal inside the upper layer of structure. This provides the evidence that transformed structure is not dense and other materials can penetrate it easily. High magnification *TEM* image of the cross-section further confirms that the transformed structure does not have any crystallinity (image shown in supporting information figure S5e).

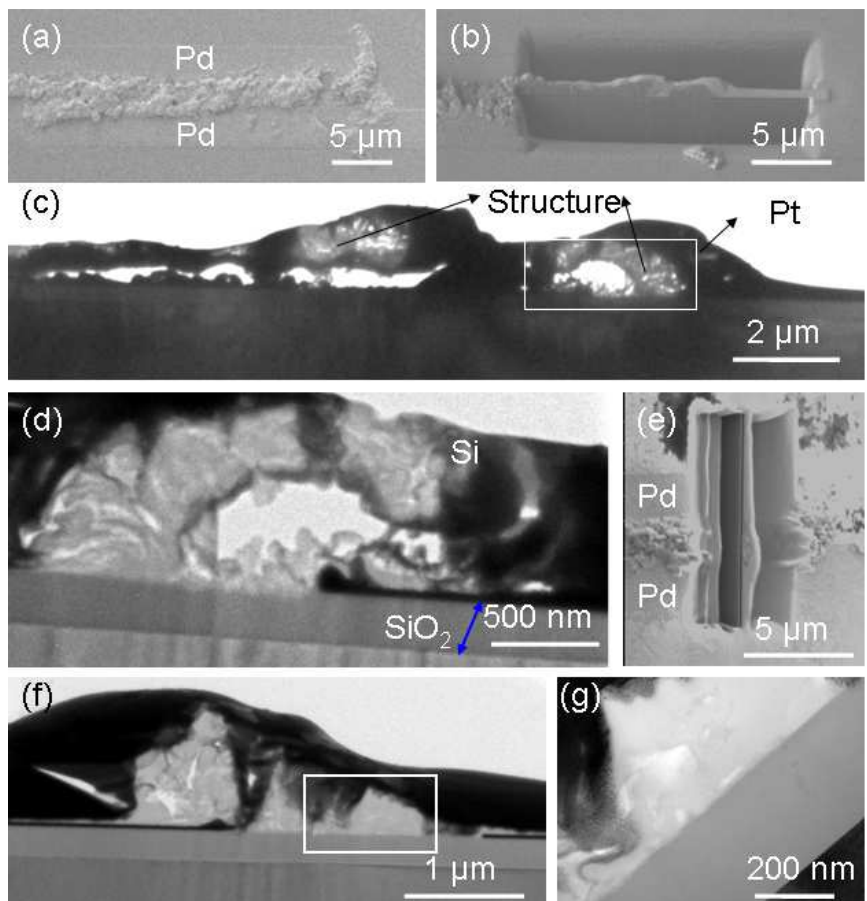

Figure 5. Cross sectional analysis by FIB and *TEM*. (a) *FIB* image of a sample. (b) Material is milled away by *Ga* ion. A thin bar (specimen) along the channel is prepared for further analysis. (c) Low resolution *TEM* image showing huge change in the structure with a large volume gain. In many places, large voids are observed (d) *HRTEM* image showing 5-10 *nm* of substrate damage at some places. (e) Another specimen is prepared orthogonal to the electrode by *FIB*. (f) *TEM* image showing similar volume expansion. (g) *HRTEM* image revealing minor damage to the substrate.

We also performed *EELS* measurement on the specimens for elemental analysis which provides information about the chemical bonding between the atoms as well. We mapped Carbon (*C*), Oxygen (*O*), and Silicon (*Si*) to estimate the presence of different elements in the mushroomed structure. Figure 6a shows a composite image for a part of specimen *B* where *C*, *O*, and *Si* atoms are mapped by blue, green, and red color respectively. The off green strip in figure 6a is $SiO_2$ whereas the red triangle is *Si* substrate. The separate images for each element are shown in the supporting information (figure S6). From these images, we conclude that transformed material consists of *C* atoms only.

Figure 6b shows the cross-loss spectra in the carbon *K*-edge region at ~284 and ~ 293.5 *eV*. A broader peak at 320 *eV* is also observed. The loss peak at 293.5 *eV* can be designated to

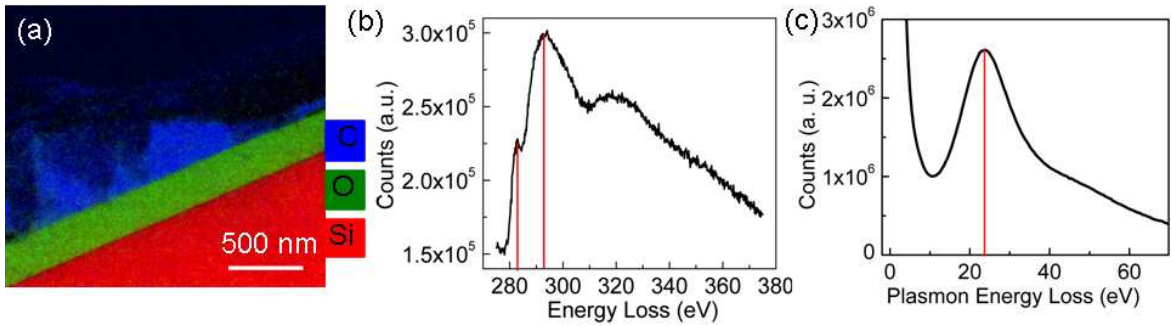

Figure 6. Elemental and structural analysis of the specimen by *EELS*. (a) Composite energy-filtered *TEM* mapping of *C*, *O* and *Si* atoms (b) *EELS* also revealing presence of *C* atoms with predominantly sp$^3$ bonding. (c) Plasmon energy loss showing a peak at 23 eV.

sp$^3$ network 1s→ σ* transition as observed in bulk diamond (~ 290 *eV*) whereas the transition peak 1s→ π* at ~ 284 *eV* is characteristic of sp$^2$ network (bulk graphite, *SWNT*, carbon onion).[46-53] We observed a very weak 1s→ π* transition peak that is due to non-availability of graphitic network (or π electron) after the transformation.[52] The intensity of 293.5 *eV* is much stronger than 284 *eV* peak providing evidence that the new structure is predominantly sp$^3$. The peak at 320 *eV* also arises due to 1s→ σ* transition and prominent in diamond related structure.[48] Figure 6c shows the plasmon-loss peak at ~ 24 *eV*. The collective excitations of the valence electrons (plasmons) in graphite are expected to have two prominent energy loss peaks at 7 and 27 *eV* respectively.[48,53] The 7 *eV* peak corresponds to π orbital whereas peak at 27 *eV* originates due to the collective excitations of three strong σ bonds and one π bond.[48,53] However, in *a-C*, the π peak is absent due to the lack of long range ordering and only one peak is observed at ~ 25 *eV*.[53] Therefore our plasmon peak at ~ 24 *eV* further confirms that the material is amorphous in nature.

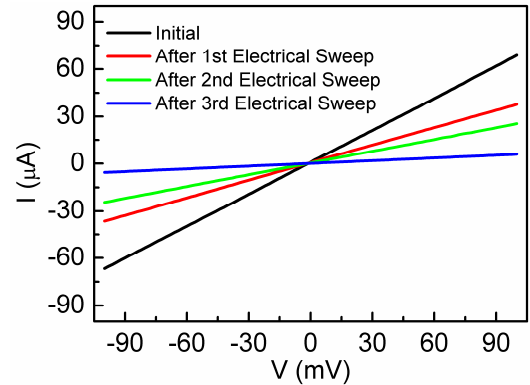

Figure 7. Electrical transport properties during field sweeping. *I-V* curves before and after the field sweeping.

Figure 7 shows the low bias *I-V* characteristics of the as fabricated device as well as after each successive voltage sweeps. At low bias *I-V* curves show Ohmic behavior. The initial resistance (*R*) of the device was 1.5 *kΩ* as calculated from the slope of the *I-V* curve. The *R* calculated from the *I-V* curves after subsequent field sweeping shows that *R* increased to 2.7 *kΩ* after the first sweep, 4 *kΩ* after the second sweep and 14.2 *kΩ* after the third sweep. This possibly indicates that sp$^3$ fraction increases after each sweep. All the information presented here indicates that *SWNT* array transformed into amorphous structure with a huge volume gain during the vacuum breakdown.

The reconstruction into new morphology requires bond breakage and rearrangements of several carbon atoms. The bond breaking in *SWNT* initiates above 1300 $^o$*C*.[54-55] Defects, dislocations, and curvature-induced strain weaken the bonds and bond rearrangement can initiate at those sites due to increased instability in the graphitic (sp$^2$) structure. We hypothesize that structural discontinuity point is the defect site where structural transformation first initiated during field sweeping and we observe maximum change at that point due to local heating. It has been reported that during joule heating, parts of the nanotubes closer to positive biased electrode get hotter.[56] Therefore we believe, the part of nanotubes closer to source (positive bias) gets swollen because of preferential heating near to the source electrode. Once the nanotube heats up locally bonds start to break and they reorganize. During the reorganization of the bonds, nanotubes appear swollen.[57] In the bond breaking process nanotubes also coalesce with neighboring tubes as observed previously.[20-21] The swollen tube which has many defects after bonds reorganization, fragmented into *a-C* particles on repeated field sweeping which provides the required energy for disintegration. In the air, nanotubes oxidizes at (600 $^oC$),[58-59] and heat dissipates to the ambient and thus prohibits any structural transformation.

The temperature rise during joule heating may provide the required energy and temperature for reconstruction. The temperature can be estimated from *I-V* curve by using heat continuity equation $\kappa A T = -IV$.[9,60] For a cylindrical rod clamped at both ends with fixed temperature giving $\Delta T = lP/(8\kappa A)$,[9,60] where *l* is nanotube length, *A* is cross sectional area, $\kappa$ is the thermal conductivity (3500 *W/mK*),[5] and *P* is the dissipated power in nanotube. The maximum current ($I_{max}$) obtained during the sweeping is ~ 8.8 *mA* at ~ 12 *V*. There are approximately 750 nanotubes in the 25 *μm* channel width hence ~11.7 *μA* current passes through each individual nanotube. Using the radius of the *SWNT* to be 1 *nm* and length to be 2 *μm*, the peak power dissipated ($P_{max}=I_{max}V$) per nanotube in our assembly is ~ 140 *μWatt* which is equivalent to 3150 $^oC$. This temperature is high enough to melt the $SiO_2$ substrate and *Pd* electrodes as the melting temperature of $SiO_2$ is 1700 $^oC$ and for *Pd* it is 1600 $^oC$. However, experimental evidences show very minimal damage to substrate at some places and no damage to *Pd* electrodes raising question whether we can use the peak power formula. On the other hand, if we use average power, we get more reasonable estimate. The average power is calculated from the area under the *I-V* curve and is found to be ~70 *μWatt* which corresponds to a temperature of ~1575 $^oC$. This temperature is slightly above of 1300 $^oC$ required for bond rearrangement but not enough for significant damage to the substrate and electrode.

## 4. Conclusions

To conclude, the aligned arrays of *SWNT* undergo a structural transformation during high field sweep in vacuum that leads to reorganization of bonds and unprecedented enhancement (~ 400 %) of volume. During electrical field sweep *SWNTs* first swells due to bond breakage and coalescence. Later they fragment into *a-C* on further electrical sweeping. The graphitic $sp^2$ network of *SWNT* is predominantly transformed to $sp^3$ during the transformation that results in an increase of the resistance of the sample. The Joule heating provides the necessary energy and temperature for the bond reorganization.

*Acknowledgments.* This work is partially supported by the U. S. National Science Foundation under Grant ECCS-0748091 (CAREER). We would like to thank Biddut K. Sarker and Abrar H. Quadery for useful discussions.

## References

1. Saito, R.; Dresselhaus, G.; Dresselhaus, M.S. Physical Properties of Carbon Nanotubes; World Scientific: Singapore, 1998.
2. Avouris, P.; Chen, Z.; Perebeinos, V. Carbon-based electronics. *Nat. Nanotechnol.* **2007**, 2, 605-615.
3. Iijima, S.; Barbec, C.; Maiti, A.; Bernholc, J. Structural Flexibility of Carbon Nanotubes. *J. Chem. Phys.* **1996**, 104, 2089-2092.
4. Javey, A.; Guo, J.; Wang, Q.; Lundstrom, M.; Dai, H. Ballistic Carbon Nanotube Field-Effect Transistor. Nature **2003**, 424, 654–657.
5. Stokes, P; Khondaker, S. I. High Quality Solution Processed Carbon Nanotube Transistor Assembled by Dielectrophoresis. *Appl. Phys. Lett.* **2010**, 96, 083110.
6. Pop, E.; Mann, D.; Wang, Q.; Goodson, K.; Dai, H. Thermal Conductance of an Individual Single-Wall Carbon Nanotube above Room Temperature. *Nano Lett.* **2006**, 6, 96-100.
7. Peng, B.; Locascio, M; Zapol, P.; Li, S.; Mielke S. L.; Schatz G. C.; Espinosa, H. D. Measurements of near-ultimate strength for multiwalled carbon nanotubes and irradiation-induced crosslinking improvements. *Nat. Nanotechnol*. **2008**, 3, 626-631.
8. Dekker, C. Carbon nanotubes as molecular quantum wires. *Physics Today* **1999**, 52, 22-28.
9. Collins, P. G.; Hersam, M.; Arnold, M.; Martel, R.; Avouris, P. Current saturation and electrical breakdown in multiwalled carbon nanotubes. *Phys. Rev. Lett.* **2001**, 86, 3128-3131.
10. Hersam, M. C. Progress towards monodisperse single-walled carbon nanotubes. Nat. Nanotechnol. **2008**, 3, 387-394.
11. Avouris, P.; Freitag, M.; Perebeinos, V. Carbon-nanotube photonics and optoelectronics. Nat. Photon. **2008**, 2, 341-350.
12. Tans, S. J.; Devoret, M. H.; Dai, H.; Thess, A.; Smalley, R. E.; Geerligs, L. J.; Dekker, Cees, Individual single-wall carbon nanotubes as quantum wires. *Nature* **1997**, 386, 474-477.
13. Stokes, P; Khondaker, S. I. Controlled fabrication of single electron transistors from single-walled carbon nanotubes. *Appl. Phys. Lett.* **2008**, 92, 262107.
14. Stokes, P; Khondaker, S. I. Evaluating Defects in Solution-Processed Carbon Nanotube Devices via Low-Temperature Transport Spectroscopy. *ACS Nano.* **2008**, 4, 2659-2666.
15. Sahoo, S.; Kontos, T.; Furer, J.; Hoffmann, C.; Graber, M.; Cottet, A.; Schonenberger, C. Electric field control of spin transport. Nat. Phys. **2005**, 1, 99-102.
16. Zhu, W.; Bower, C.; Zhou, O.; Kochanski, G.; Jin, S. Large current density from carbon nanotube field emitters. App. Phys. Lett. **1999**, 75, 873-875.
17. Dujardin, E.; Ebbesen, T. W.; Hiura, H.; Tanigaki, K. Capillarity and Wetting of Carbon Nanotubes. *Science* **1994**, 265, 1850-1852.
18. Huang, J.Y.; Chen, S.; Wang, Z.Q.; Kempa, K.; Wang, Y.M.; Jo, S. H.; Chen, G.; Dresselhauss, M.S.; Ren, Z. F. Superplastic carbon nanotubes. *Nature* **2006**, 439, 281.
19. Jeong, S. J.; Park, K. A.; Jeong, S. H.; Jeong, H. J.; An, K. H.; Nah, C. W.; Pribat, D.; Lee, S. H.; Lee, Electroactive Superelongation of Carbon Nanotube Aggregates in Liquid Crystal Medium. *Nano Letters* **2007**, 7, 2178-2182.

20. Nikolaev, P.; Thess, A.; Rinzler, A. G.; Colbert, D. T.; Smalley, R. E. Diameter doubling of single-wall nanotubes. *Chem. Phys. Lett.* **1997**, 266, 422-426.
21. Terrones, M.; Terrones, H.; Banhart, F.; Charlier, J. C.; Ajayan, P. M. Coalescence of Single-Walled Carbon Nanotubes. *Science* **2000**, 288, 1226-1229.
22. Krasheninnikov, A.V.; Banhart, F. Engineering of nanostructured carbon materials with electron or ion beams. Nat. Mater. **2007**, 6, 723-733.
23. Zhang, M., Fang, S.; Zakhidov, A. A.; Lee, S. B.; Aliev, A. E.; Williams, C. D.; Atkinson, K. R.; Baughman, R. H. Strong, Transparent, Multifunctional, Carbon Nanotube Sheets. *Science*, **2005**, 309, 1215-1219.
24. Rutherglen, C.; Jain, D.; Burke, P. Nanotube Electronics for Radiofrequency Applications. *Nat. Nanotechnol.* **2009**, 4, 811-819.
25. Shekhar, S.; Stokes, P.; Khondaker, S. I. Ultrahigh Density Alignment of Carbon Nanotube Arrays by Dielectrophoresis. *ACS Nano*, **2011**, 5, 1739-1746.
26. Stokes, P; Khondaker, S. I. Solution Processed Large Area Field Effect Transistors from Dielectrophoretically Aligned Arrays of Carbon Nanotubes. *Appl. Phys. Lett.* **2009**, 94, 113104.
27. Sarker, B. K.; Liu, J.; Zhai, L.; Khondaker, S. I. Fabrication of Organic Field Effect Transistor by Directly Grown Poly(3 Hexylthiophene) Crystalline Nanowires on Carbon Nanotube Aligned Array Electrode. *ACS Applied Materials & Interfaces* **2011**, 3, 1180-1185.
28. Kocabas, C.; Kim, H.-s.; Banks, T.; Rogers, J.A.; Pesetski, A.A.; Baumgardner, J.E.; Krishnaswamy, S.V.; Zhang H. Radio Frequency Analog Electronics Based on Carbon Nanotube Transistors. *Proc. Natl. Acad. Sci. U.S.A.* **2008**, 105, 1405-1409.
29. Rutherglen, C.; Jain, D.; Burke, P. rf Resistance and Inductance of Massively Parallel Single Walled Carbon Nanotubes: Direct, Broadband Measurements and Near Perfect 50 Ω Impedance Matching. *Appl. Phys. Lett.* **2008**, 93, 083119-1-083119-3.
30. Pesetski, A. A.; Baumgardner, J. E.; Krishnaswamy, S. V.; Zhang, H.; Adam, J. D.; Kocabas, C.; Banks, T.; Rogers, J. A. A 500 MHz Carbon Nanotube Transistor Oscillator. *Appl. Phys. Lett.* **2008**, 93, 123506-1-123506-3.
31. Kocabas, C.; Dunham, S.; Cao, Q.; Cimino, K.; Ho, X.; Kim, H.-S.; Dawson, D.; Payne J.; Stuenkel, M.; Zhang, H.; Banks, T.; Feng, M.; Rotkin, S.V.; Rogers, J. A. High-Frequency Performance of Submicrometer Transistors That Use Aligned Arrays of Single-Walled Carbon Nanotubes. *Nano Lett.* **2009**, 9, 1937-1943.
32. Kang, S. J.; Kocabas, C.; Ozel, T.; Shim, M.; Pimparkar, N.; Alam, M. A.; Rotkin, S. V.; Rogers, J. A. High-Performance Electronics Using Dense, Perfectly Aligned Arrays of Single-Walled Carbon Nanotubes. *Nat. Nanotechnol.* **2007**, 2, 230-236.
33. Kocabas, C.; Hur, S.-H.; Gaur, A.; Meitl, M.A.; Shim, M.; Rogers, J.A. Guided Growth of Large-Scale, Horizontally Aligned Arrays of Single-Walled Carbon Nanotubes and Their Use in Thin-Film Transistors. *Small* **2005**, 1, 1110-1116.
34. Ryu, K.; Badmaev, A.; Wang, C.; Lin, A.; Patil, N.; Gomez, L.; Kumar, A.; Mitra, S.; Wong, H.-S. P.; Zhou, C. CMOS-Analogous Wafer-Scale Nanotube-on-Insulator Approach for Submicrometer Devices and Integrated Circuits Using Aligned Nanotubes. *Nano Lett.* **2009**, 9, 189-197.
35. McNicholas, T. P.; Ding, L.; Yuan, D.; Liu, J. Density Enhancement of Aligned Single-Walled Carbon Nanotube Thin Films on Quartz Substrates by Sulfur-Assisted Synthesis. *Nano Lett.* **2009**, 9, 3646-3650.

36. Hong S.W.; Banks T.; Rogers J. A. Improved Density in Aligned Arrays of Single-Walled Carbon Nanotubes by Sequential Chemical Vapor Deposition on Quartz. *Adv. Mater.* **2010,** 22, 1826-1830.
37. Ishikawa, F. N.; Chang, H.-k.; Ryu, K.; Chen, P.-c.; Badmaev, A.; De Arco, L.G.; Shen, G.; Zhou, C. Transparent Electronics Based on Transfer Printed Aligned Carbon Nanotubes on Rigid and Flexible Substrates. *ACS Nano* **2008**, 3, 73-79.
38. Cao, Q.; Rogers, J. A. Ultrathin Films of Single-Walled Carbon Nanotubes for Electronics and Sensors: A Review of Fundamental and Applied Aspects. *Adv. Mater.* **2009**, 21, 29-53.
39. Kim, S.; Ju, S.; Back, J. H.; Xuan, Y.; Ye, P. D.; Shim, M.; Janes, D. B.; Mohammadi, S. Fully Transparent Thin-Film Transistors Based on Aligned Carbon Nanotube Arrays and Indium Tin Oxide Electrodes. *Adv. Mater.* **2009**, 21, 564-568.
40. Zhou, W.; Rutherglen, C.; Burke, P. J. Wafer Scale Synthesis of Dense Aligned Arrays of Single-Walled Carbon Nanotubes. *Nano Res.* **2008**, 1, 158-165.
41. Shekhar, S.; Erementchouk, M.; Leuenberger, M. N.; Khondaker, S. I. Correlated electrical breakdown in arrays of high density aligned carbon nanotubes. *Appl. Phys. Lett.*, **2011**, 98, 243121.
42. Brewer Science Inc. http://www.brewerscience.com/products/carbon-nanotube/ (Accessed May 24, 2011).
43. Lucas, A. A.; Moreau, F.; Lambin P. Optical simulations of electron diffraction by carbon nanotubes. *Rev. Mod. Phys.* **2002**, 74,1-10.
44. Ajayan, P. M. Nanotubes from Carbon. *Chem. Rev.* **1999**, 99 1787-1800.
45. Liao, A.; Alizadegan, R.; Ong, Z.-Y.; Dutta, S.; Xiong, F.; Hsia, K. J.; Pop, E. Thermal dissipation and variability in electrical breakdown of carbon nanotube devices. *Phys. Rev. B* **2010**, 82, 205406.
46. Knupfer, M.; Pichler, T.; Golden, M. S.; Fink, J.; Rinzler, A.; Smalley, R. E. Electron Energy-Loss Spectroscopy Studies of Single Wall Carbon Nanotubes. *Carbon* **1999**, 37, 733-738.
47. Tomita, S.; Fujii, M.; Hayashi, S.; Yamamoto, K. Electron energy-loss spectroscopy of carbon onions. *Chem. Phys. Lett.* **1999**, 305, 225-229.
48. Castrucii, P.; Scarselli, M.; Crescenzi, D.; Maurizio K.; My Ali El, Probing the electronic structure of carbon nanotubes by nanoscale spectroscopy. *Nanoscale* **2010**, 2, 1611-1625.
49. Stephan, O.; Kociak, M.; Henrard, L.; Suenaga, K; Gloter, A.; Tence, M.; Sandre, E.; Colliex, C. Electron energy-loss spectroscopy on individual nanotubes. *J. Elec. Spectroscopy and Related Phenomena* **2001**, 114-116, 209-217.
50. Sun, X.-H.; Li, C.-P.; Wong, W.-K.; Wong, N.-B.; Lee, C.-S.; Lee, S.-T.; Teo, B.-K. Formation of Silicon Carbide Nanotubes and Nanowires via Reaction of Silicon (from Disproportionation of Silicon Monoxide) with Carbon Nanotubes. *J. Am. Chem. Soc.* **2002**, 124, 14464-14471.
51. Ferrari, A. C.; Libassi, A.; Tanner, B. K.; Stolojan, V.; Yuan, J.; Brown, L. M.; Rodil, S. E.; Kleinsorge, B.; Robertson, J. Density, $sp^3$ fraction, and cross-sectional structure of amorphous carbon films determined by x-ray reflectivity and electron energy-loss spectroscopy. *Phys. Rev. B* **2000**, 62, 11089.
52. Gago, R.; Vinnichenko, M. A.; Jager, H. U.; Belov, A. Y.; Jimenez, I.; Huang, N.; Sun, H.; Maitz, M. F. Evolution of $sp^2$ networks with substrate temperature in amorphous carbon films: Experiment and theory. *Phys. Rev. B* **2005**, 72, 014120.

53. Ajayan, P.M.; Iijima, S.; Ichihashi, T. Electron-energy-loss spectroscopy of carbon nanometer-size tubes. *Phys. Rev. B* **1993**, 47, 6859.
54. Ajayan, P. M.; Terrones, M.; de la Guardia, A.; Huc, V.; Grobert, N.; Wei, B. Q.; Lezec, H.; Ramanath, G.; Ebbesen, T. W. Nanotubes in a Flash-Ignition and Reconstruction. *Science* **2002**, 296, 705.
55. Gutierrez, H. R.; Kim, U. J.; Kim, J. P.; Eklund, P. C. Thermal Conversion of Bundled Carbon Nanotubes into Graphitic Ribbons. *Nano Letters*, **2005**, 5, 2195-2201.
56. Estrada, D.; Pop E. Imaging dissipation and hot spots in carbon nanotube network transistors. *Appl. Phys. Lett.* **2011**, 98, 073102.
57. Yu, M. -F.; Dyer, M. J.; Ruoff, R. S. Structure and mechanical flexibility of carbon nanotube ribbons: An atomic-force microscopy study. *J. Appl. Phys.* **2001**, 89, 4554-4557.
58. Hata, K.; Futaba, D. N.; Mizuno, K.; Namai, T.; Yumura, M.; Iijima, S. Water-Assisted Highly Efficient Synthesis of Impurity-Free Single-Walled Carbon Nanotubes. *Science* **2004**, 306, 1362-1364.
59. Eric, P. The role of electrical and thermal contact resistance for Joule breakdown of single-wall carbon nanotubes. *Nanotechnology* **2008**, 19, 295202.
60. Molhave, K.; Gudnason, S. B.; Pedersen, A. T.; Clausen, C. H.; Horsewell, A.; Boggild, P. Transmission Electron Microscopy Study of Individual Carbon Nanotube Breakdown Caused by Joule Heating in Air. *Nano Lett.* **2006**, 6, 1663-1668.

# Supporting Information

**Huge Volume Expansion and Structural Transformation of Carbon Nanotube Aligned Arrays during Electrical Breakdown in Vacuum**

Shashank Shekhar [1,2], Helge Heinrich [2,3], and Saiful I. Khondaker [1,2,4]*

[1] Nanoscience Technology Center, [2] Department of Physics, [3] Advanced Material Processing and Analysis Center, [4] School of Electrical Engineering and Computer Science, University of Central Florida, Orlando, Florida 32826, USA.

**Length and diameter distribution of carbon nanotubes in the solution:**

Figure S1a shows a scanning electron micrograph (*SEM*) of nanotubes dispersed on $SiO_2$ substrate by spin coating. The spin speed was 1000 *rpm*. Interestingly, the dispersed nanotubes are also aligned during the spin coating. From many images like this one, we calculated the length of the nanotubes. The length distribution is shown in figure S1b from ~ 150 nanotubes. We found a mean length of ~ 1.5 *μm*. For diameter study, we spun the nanotubes on a mica substrate and took atomic force microscopy (*AFM*) images. Figure S1c shows an *AFM* image of such a nanotube with height profile shown in figure S1d. The diameter distribution from many such measurements is shown in figure S6e. The average diameter is ~1.7 *nm*. From this, we conclude that the solution contains mostly individual nanotubes and are free from catalytic particle.

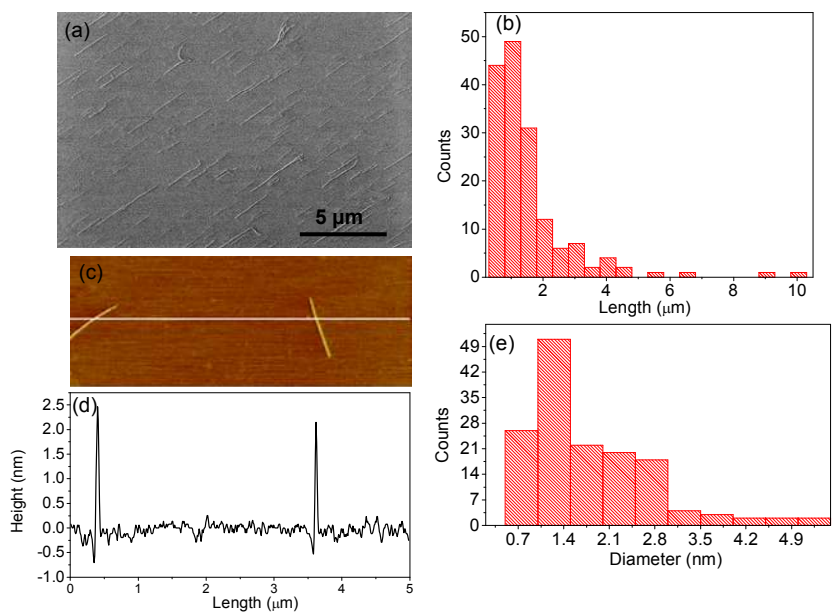

Figure S1. (a) *SEM* image of *SWNTs* dispersed via spin coating on a $SiO_2$ substrate. From images like this, we obtained length distribution of nanotubes shown in (b) Average length is ~ 1.5 *μm*. (c) *AFM* image of *SWNTs* along with height profile shown in (d). (e) Diameter distribution of the *SWNTs* in the solution based on *AFM* height measurements. Average diameter is ~ 1.7 *nm*.

**Optical image of aligned arrays of SWNTS:**

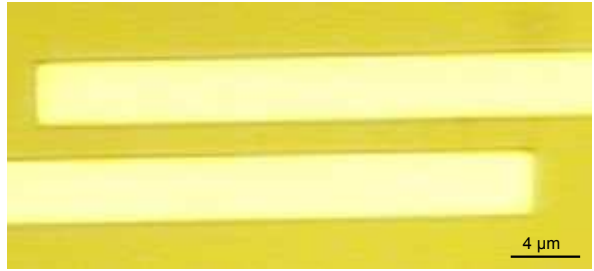

Figure S2. Optical image of aligned array of *SWNTs*. Being transparent to visible light nanotubes are not seen

Figure S2 shows the optical image of aligned arrays of *SWNT* on a device of channel length 2 *μm* and channel width 25 *μm*. The aligned arrays are optically transparent to the optical wavelengths and we see substrate in the image. The *SEM* image (figure 1b) shows that the electrodes are densely bridged by nanotubes.

**AFM analysis of partially swollen tube during controlled breakdown:**

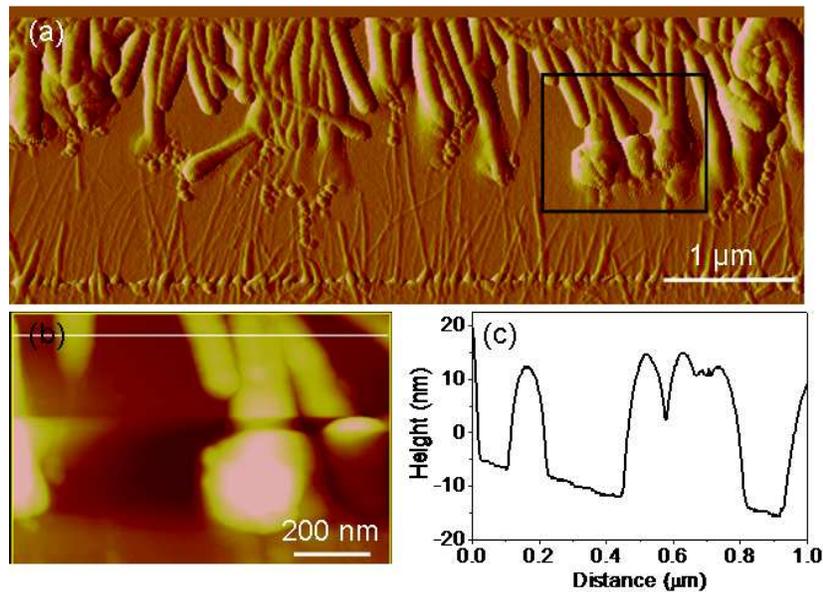

Figure S3. *AFM* images of transformed *SWNT* after controlled electrical sweeping. (a) Amplitude image of transformed tubes (b) Height image of designated area (c) Height analysis shows that diameter of the tubes to be 20-30 *nm*.

Figure S3 is *AFM* analysis of transformed *SWNTs* aligned arrays after partial electrical sweep. Figure S3a shows the amplitude image of the aligned array after controlled electrical field sweep of 10 *V/μm* (figure 4a). Figure S3b is the high magnification height image of *AFM* of swollen tubes. Figure S3c is the corresponding height analysis. The diameter of the measured tube is 20-30 *nm*. The bulbs seen in the images are much larger in diameter and sometimes their diameters are as big as 200 *nm*.

**Additional SEM images of amorphous carbon of varying shape and size:**

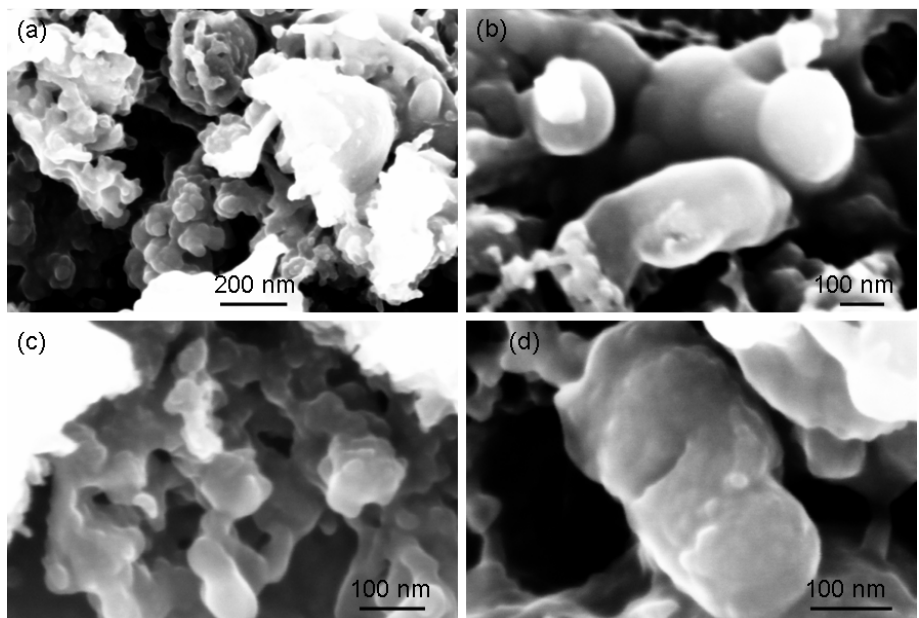

Figure S4. *SEM* images show amorphous carbon with different shapes and sizes

*SEM* images (figure S4) show amorphous carbon particles of different shape and size. Few of them appears to be as big as 500 *nm* (seen in the figure S4a) due to merging of smaller particles.

***TEM* image of interface between amorphous carbon-substrate:**

Figurse S5 shows the *HRTEM* image of amorphous carbon-substrate interface. The *HRTEM* images (figure S5a and S5b) show the close view of the interface. The roughness of the interface is of the order of 10 *nm* which may be due to substrate damage during the electrical field sweeping. The diffusion of *Pt* is also visible in the *HRTEM* images. Figure S5c shows the diffusion profile of *Pt* metal. At some places *Pt* penetrates deep inside and comes very close to the substrate. The high magnification image (figure S5d) shows the penetration of sub 10 *nm* Pt particles. Figure S5e is the high magnification image of the material (away from *Pt* diffusion) shows amorphous nature as discussed in main text.

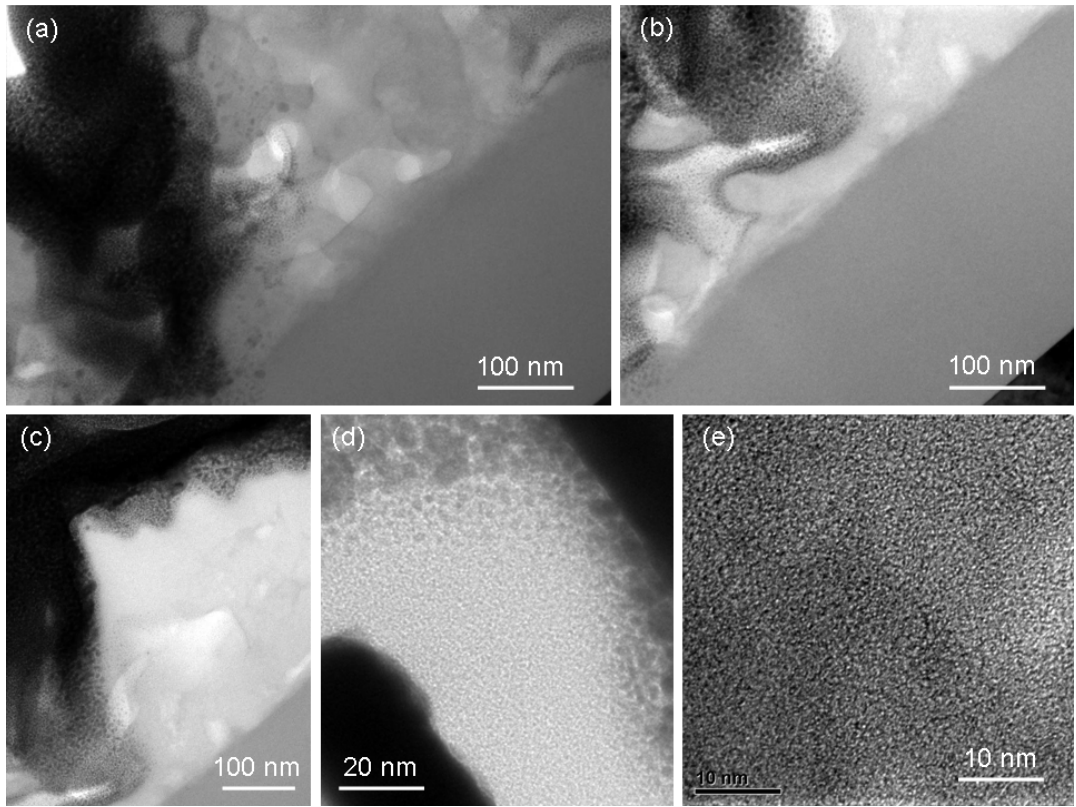

Figure S5. High resolution *TEM* image of substrate-amorphous carbon interface (a,b) show the interface between *SiO₂* and transformed structure. The interface shows the profile with 5-10 *nm* of roughness. That may be due to substrate damage. The black dots are platinum diffusion in the structure. (c) Low magnification image shows the Pt diffuses sometimes down to the substrate (d) High magnification image shows diffusion of sub 10 *nm* Pt particles. (e) Amorphous structure is seen in the high resolution image.

**Elemental mapping by *EELS* spectroscopy:**

Figure S6 shows the elemental analysis of individual element by *EELS* spectroscopy. Figure S6a is the *HRTEM* image of the sample which is used for elemental mapping. *Si/SiO₂* substrate is seen clearly in figure S6a. Above substrate mushroomed structure is formed. The Platinum metal is deposited on the mushroomed structure before milling. Figure S6b shows the Carbon (*C*) atom mapping. We monitored *C*-signal from mushroomed structure only. Figure S6c is the elemental mapping of oxygen atom. Image shows oxygen (*O*) signal from *SiO₂* substrate only. Similarly we observed *Si* signal from *SiO₂/Si* substrate only (figure S6d). Figure S6e is the *Pt* mapping and it is found on the top of the mushroomed structure. At some places it has penetrated deep due to porosity in the structure. We also performed Gallium (*Ga*) mapping to confirm that the specimen did not get contaminated during milling process. Figure S6f is the Ga mapping that shows that mushroomed structure does not have any trace for *Ga* atoms.

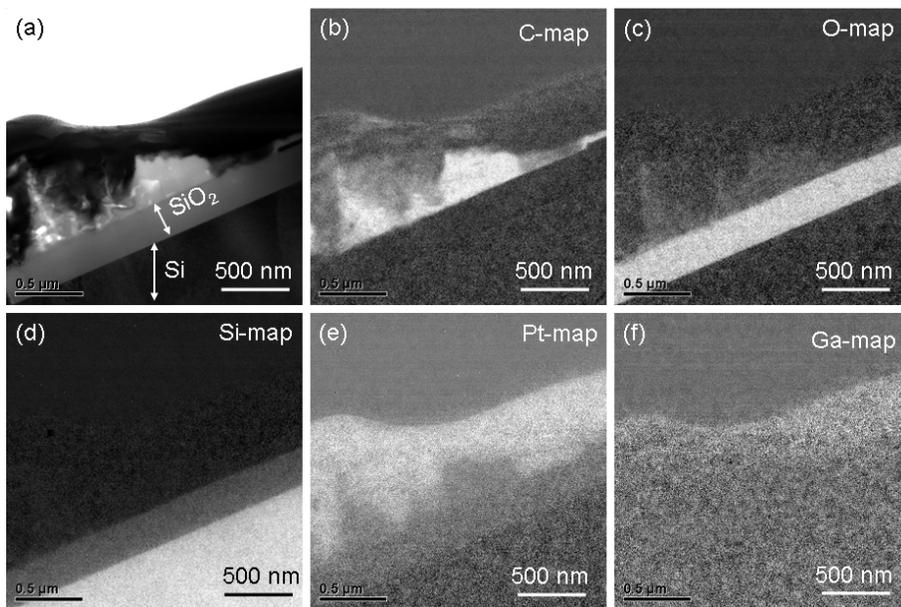

Figure S6. *EELS* elemental mapping. (a) *EELS* spectroscopy was performed on the area shown in *HRTEM* (b) The carbon mapping shows a strong signal (white color) that implicates abundance of carbon in mushroomed structure. (c & d) show Oxygen and Silicon mapping respectively. The mushroomed structure does not show noticeable signal. Both oxygen and silicon signal is found in $SiO_2$ substrate. (e) *Pt* signal is detected in upper layers of mushroomed structure. It has diffused in the mushroomed structure. (f) *Ga* mapping shows no contamination in the mushroomed structure.